\documentclass[12pt]{iopart}

\usepackage{graphicx}
\usepackage{ccaption}
\begin{document}

\title[RHIC Results on $J/\psi$]{RHIC Results on $J/\psi$}

\author{M.J. Leitch}

\address{P-25, MS H846, Los Alamos National Laboratory, Los Alamos NM 87545}
\ead{leitch@lanl.gov}
\begin{abstract}
Quarkonia ($J/\psi$, $\psi$', $\Upsilon$) production provides a sensitive probe
of gluon distributions and their modification in nuclei; and is a
leading probe of the hot-dense (deconfined) matter created in
high-energy collisions of heavy ions. We will discuss the current
understanding of the production process and of the
cold-nuclear-matter effects that modify this production in nuclei
in the context of recent p+p and
p(d)+A quarkonia measurements. Then we will review the latest results for
nucleus-nucleus collisions from RHIC, and together with the baseline
results from d+A and p+p collisions, discuss several alternative
explanations for the observed suppressions and future prospects for
distinguishing these different pictures.
\end{abstract}


\section{Introduction}
We will give an overview of the physics and the most recent measurements from
RHIC for $J/\psi$ production and suppression starting with 1) production issues as seen
in p+p collisions, then 2) cold nuclear matter effects as seen in p+A or d+A collisions,
and finally 3) effects of the hot-dense partonic matter created in heavy-ion collisions
and interpretation of the latest PHENIX heavy-ion data.
For more details on these and related topics please see other contributions to this proceedings
including those from A. Bickely (PHENIX p+p $J/\psi$), A. Glenn and T. Gunji (PHENIX A+A $J/\psi$);
P. Djawotho (STAR Upsilons and $J/\psi$); R. Granier de Cassanac and R. Vogt (cold nuclear
matter effects on $J/\psi$); E. Scomparin (NA60 $J/\psi$'s); and A. Suaide (open charm at RHIC).

\section{$J/\psi$ Production in p+p Collisions}

Gluon fusion dominates the production of quarkonia, but the configuration of the
produced state and how it hadronizes remain uncertain. Absolute cross sections
can be reproduced by NRQCD models that involve a color octet state\cite{beneke},
but these models predict transverse polarization of the $J/\psi$ at large $p_T$ which is
not seen in the data\cite{beneke2}. A complication in understanding the $J/\psi$ results is
the fact that $\sim$40\% of the $J/\psi$s come from decays of higher mass resonances
($\psi'$ and $\chi_C$)\cite{abt} - a feature that may contribute to the lack of polarization seen.

\begin{figure}[tbh]
\centering
\begin{minipage}[t]{0.42\linewidth}
  \centering
  \includegraphics[width=\textwidth,clip=]{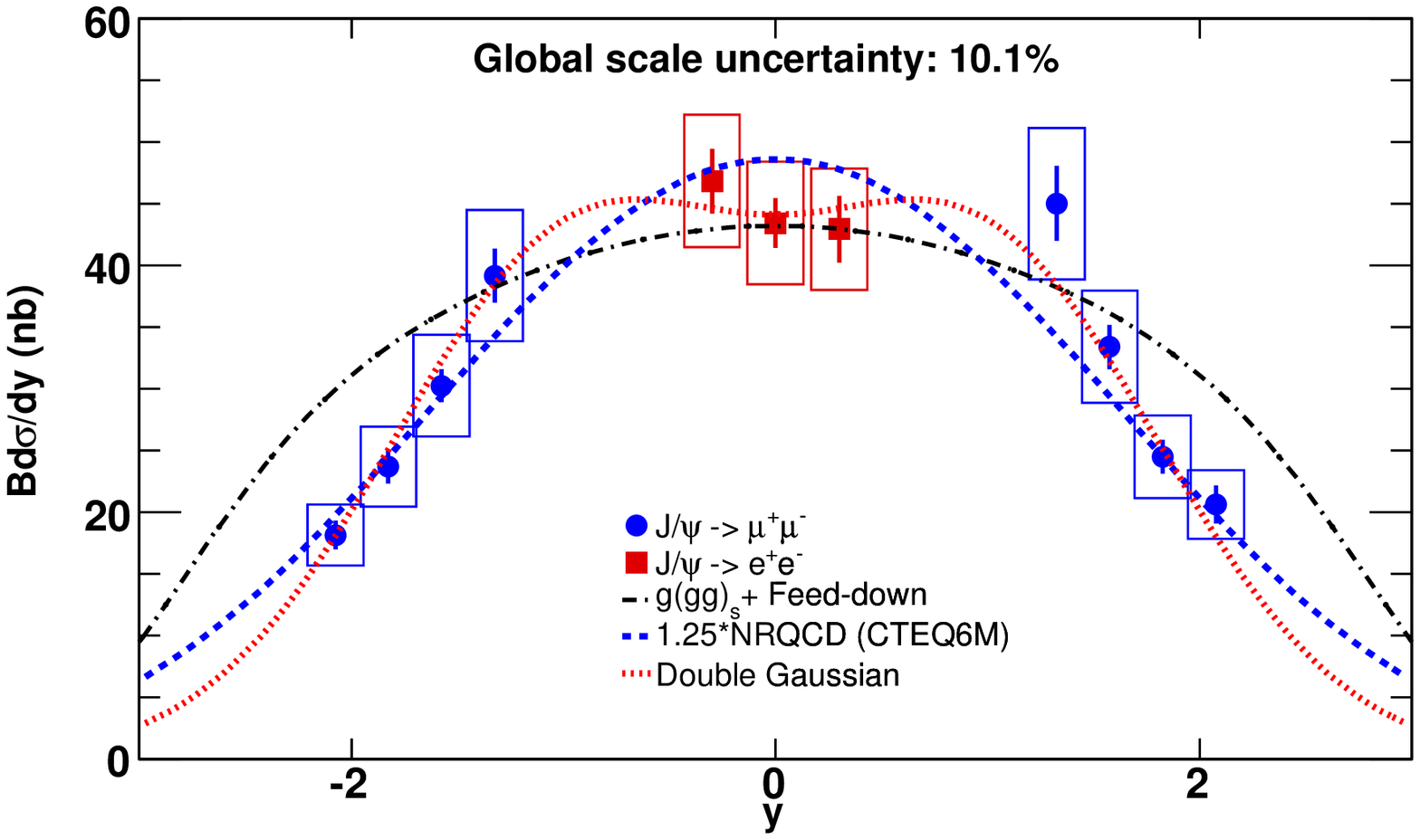}
  \caption{$J/\psi$ cross section vs rapidity for 200 GeV p+p collisions at
  RHIC\cite{phenix_jpsi_run5pp} Also shown are fits using shapes from two theoretical
  models and from a double Gaussian.}
  \label{fig:run5pp_y}
\end{minipage}
\hspace{1cm}
\begin{minipage}[t]{0.42\linewidth}
  \centering
  \includegraphics[width=\textwidth,clip=]{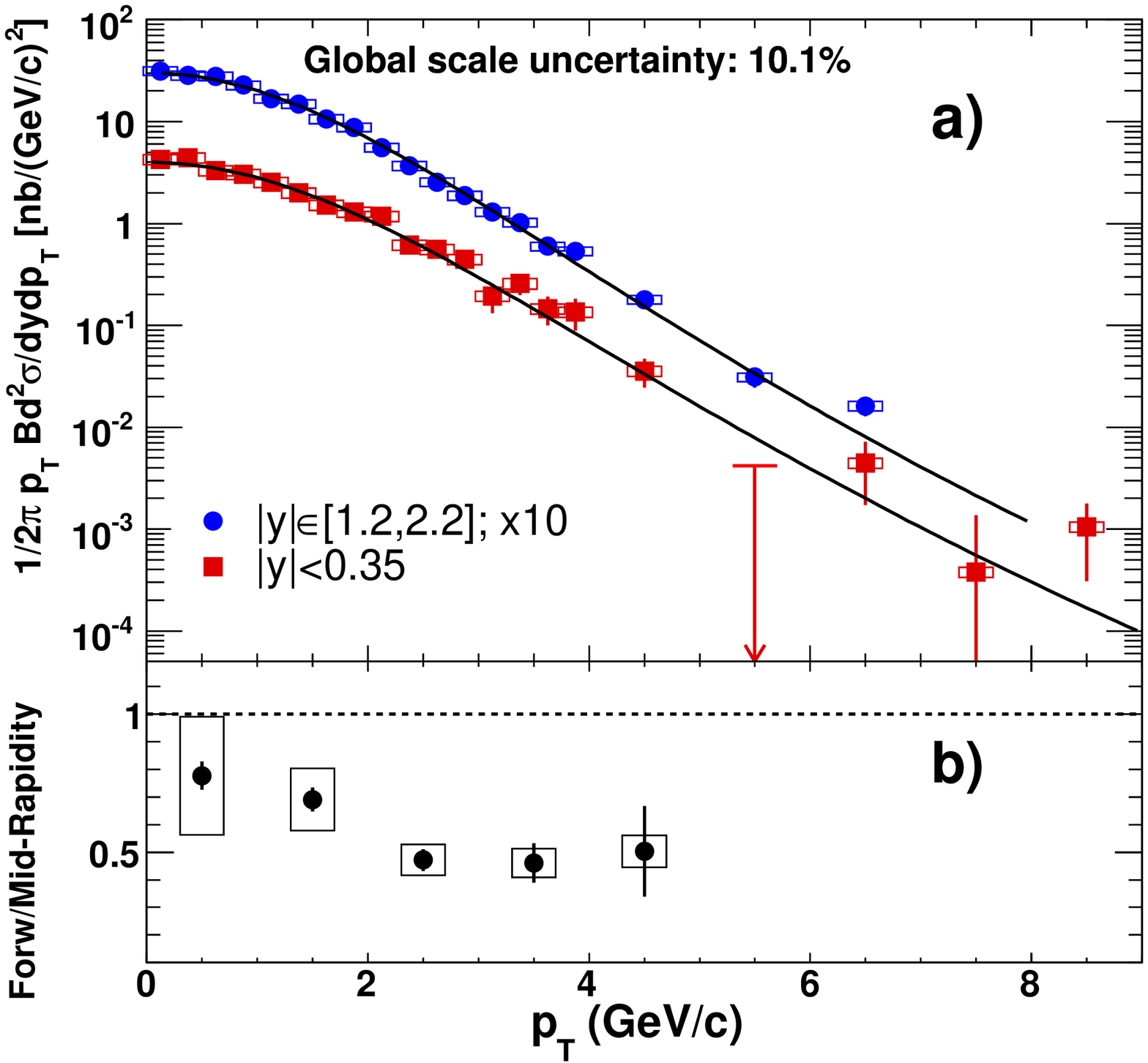}
  \caption{$J/\psi$ cross section (a) vs $p_T$ for 200 GeV p+p collisions at
  RHIC\cite{phenix_jpsi_run5pp} and (b) the ratio of the forward ($1.2<|y|<2.2$) to
  mid ($|y|<0.35$) rapidity cross sections vs $p_T$.}
  \label{fig:run5pp_pt_ratio}
\end{minipage}
\end{figure}

The most recent $J/\psi$ cross section measurements for p+p collisions at $\sqrt{s}=200$ GeV from
PHENIX\cite{phenix_jpsi_run5pp} are shown in Fig.~\ref{fig:run5pp_y}. These data slightly favor
a flatter rapidity distribution at mid rapidity than most model calculations which have shapes similar to the
NRQCD calculation (dashed curve) in the figure. A more recent pQCD calculation\cite{khoze} that includes
explicit treatment of the third gluon, necessary to give the final color singlet state,
gives good agreement with the cross sections and polarization seen in other measurements, but does not
reproduce the steep falloff at large rapidity of the PHENIX results. In Fig.~\ref{fig:run5pp_pt_ratio}
the $p_T$ distributions are also shown. The distribution is harder at mid-rapidity than
for forward rapidity with $<p_T^2> = 4.14 \pm 0.18 + 0.30 - 0.20$ (mid rapidity) and 
$3.59 \pm 0.06 \pm 0.16$ (forward rapidity), and both are harder than at lower energies.
These $<p_t^2>$ values are obtained from a fit to the data using the standard form,
$A \times (1+(p_T/B)^2)^{-6}$.

\section{Cold Nuclear Matter Effects and $J/\psi$ Suppression in p(d)+A Collisions}

\begin{figure}[tbh]
\hfill
\begin{minipage}[t]{0.36\linewidth}
  \includegraphics[width=0.9\textwidth,clip=]{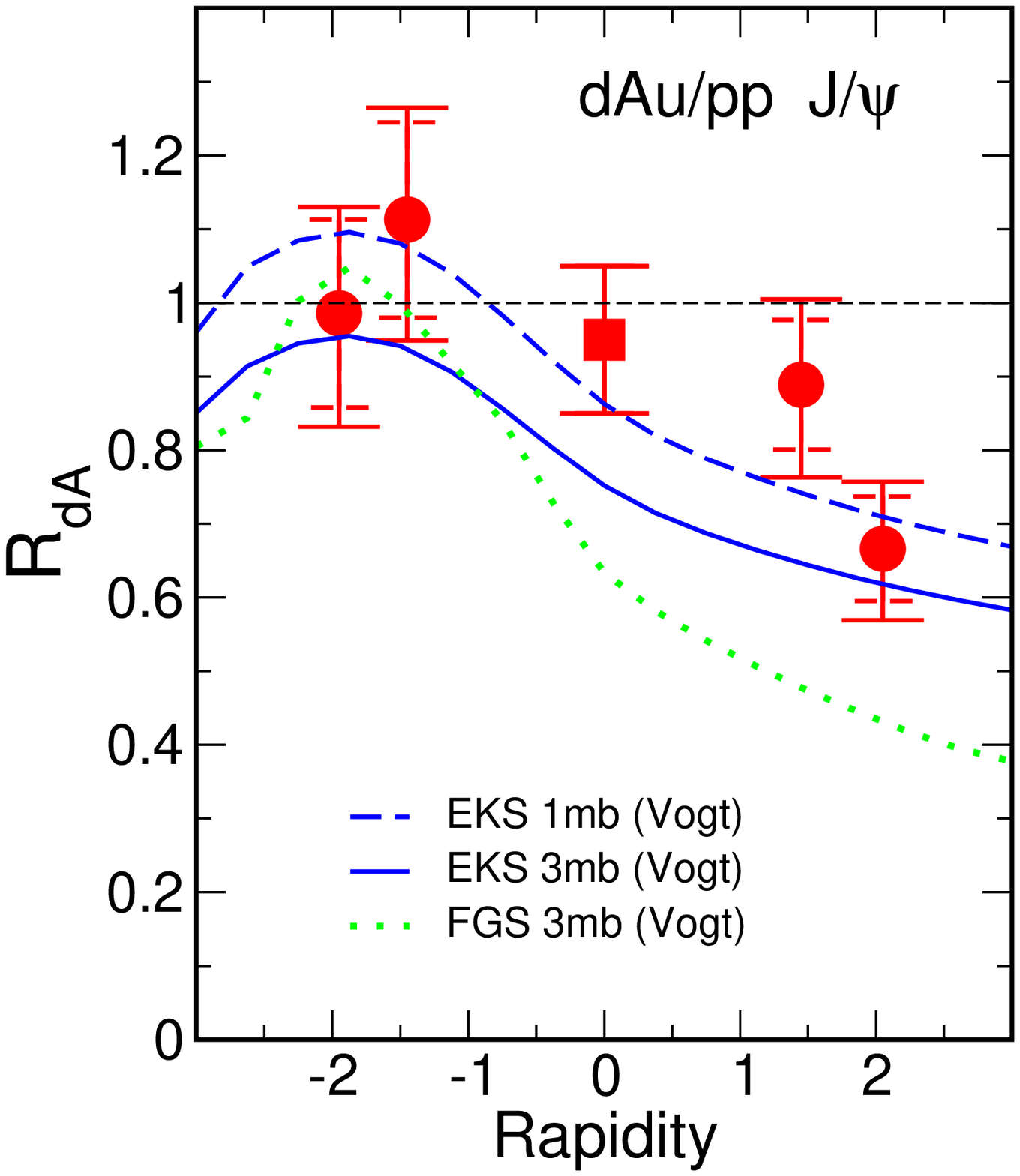}
  \caption{Rapidity dependence of the $J/\psi$ nuclear modification factor, $R_{dAu}$ for 200 GeV d+Au
  collisions at RHIC\cite{phenix_jpsi_pp}.}
  \label{fig:rda}
\end{minipage}
\hfill
\begin{minipage}[t]{0.54\linewidth}
  \includegraphics[width=\textwidth,clip=]{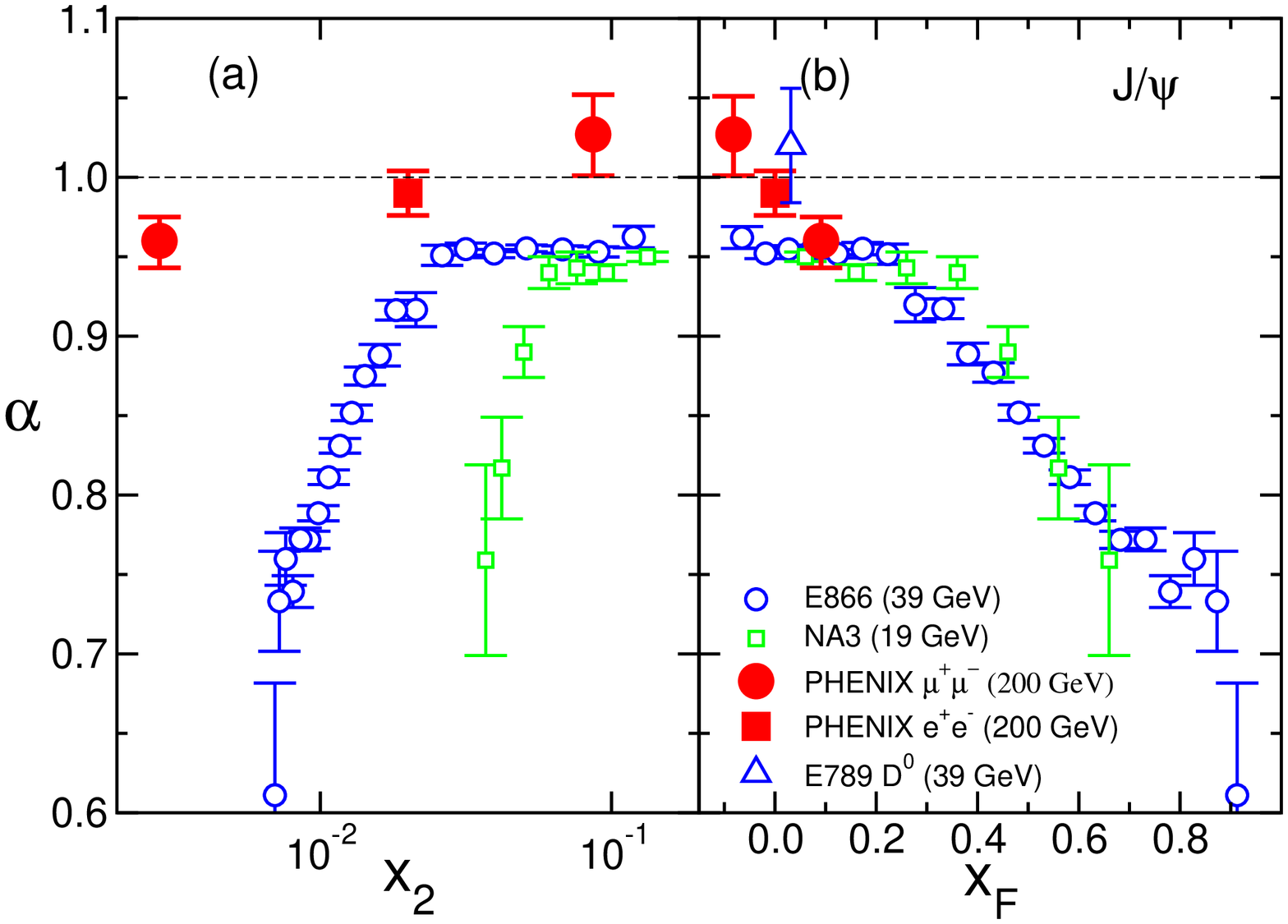}
  \caption{Test of scaling vs $x_2$ and $x_F$ for $J/\psi$ suppression data for
  three different collision energies. $J/\psi$ data is from Refs.\cite{e866jpsi,na3,phenix_jpsi_pp}
  and $D^0$ point is from Ref.\cite{e789d}.}
  \label{fig:alpha_x2xf_d}
\end{minipage}
\hfill
\end{figure}

When quarkonia are produced in nuclei their yields per nucleon-nucleon collision are known
to be significantly modified. This modification, shown vs. rapidity in Fig.~\ref{fig:rda} at
RHIC energy and vs. $x_F$ in Fig.~\ref{fig:alpha_x2xf_d}b 
is thought to be due to several cold nuclear matter (CNM) effects
including gluon shadowing, initial-state gluon energy loss and multiple scattering, and
absorption (or dissociation) of the $c\bar{c}$ in the final-state before it can form a $J/\psi$.

Shadowing is the depletion of low-momentum partons (gluons in this case) in a nucleon
embedded in a nucleus compared to their population in a free nucleon. The predicted strength of the
depletion differs between numerous models by up to a factor of three. Some models are based
on phenomenological fits to deep-inelastic scattering and Drell-Yan data\cite{eks}, while others
obtain shadowing from coherence effects in the nuclear medium\cite{fgs,boris_shadowing}.
In addition, models such as the Color Glass Condensate (CGC)\cite{cgc} obtain shadowing through
gluon saturation pictures where non-linear effects for the large gluon populations at very small x
in a nucleus generate a deficit of gluons per nucleon at small x.

In the final state, the produced $c{\bar{c}}$ can be disassociated or absorbed
on either the nucleus itself, or on light co-moving partons produced when the
projectile proton or deuteron enters the nucleus. The latter is probably only important
in nucleus-nucleus collisions as the number of co-movers created in a p+A or d+A
collisions is small.

However, $J/\psi$ suppression in p(d)+A collisions remains a puzzle given that one does not
find a universal suppression vs $x_2$ as would be expected from shadowing, Fig.~\ref{fig:alpha_x2xf_d}a; 
while vs. $x_F$ the dependence is similar for all energies, Fig.~\ref{fig:alpha_x2xf_d}b. This apparent $x_F$
scaling supports explanations such as those that involve initial-state energy loss or Sudakov
suppression\cite{sudakov}.

\section{$J/\psi$ Suppression in the Hot-dense Partonic matter created in Nucleus-Nucleus Collisions}

One of the leading predictions for the hot-dense matter created in high-energy heavy-ion
collisions was that if a deconfined state of quarks and gluons is created, i.e. a
quark-gluon plasma (QGP), the heavy-quark bound states would be screened by the
deconfined colored medium and destroyed before they could be formed\cite{matsui_satz}.
This screening
would depend on the particular heavy-quark state, with the $\psi'$ and $\chi_C$ being dissolved
first; next the $J/\psi$ and then the $\Upsilon$'s only at the highest QGP temperatures. The CERN SPS
measurements\cite{na50} showed a suppression for the $J/\psi$ and $\psi'$ beyond what was expected from CNM
effects - as represented by a simple absorption model constrained to p+A data. In
addition to explanations involving creation of a QGP, a few theoretical models\cite{capella}  were also
able to explain the data without including a QGP, so the evidence that a QGP was formed
was controversial.

\begin{figure}[tbh]
\hfill
\begin{minipage}[t]{0.58\linewidth}
  \includegraphics[width=\textwidth,clip=]{figures/fig5_aa_cent_eks_band.eps}
  \caption{$J/\psi$ suppression in Au+Au\cite{phenix_jpsi_run4auau} and Cu+Cu\cite{phenix_qm05_jpsi}
  collisions for forward rapdidity
  and central rapdity\cite{phenix_qm05_jpsi} compared to predictions for CNM from the same calculations
  as shown in Fig.~\ref{fig:rdau_eks98_band}\cite{vogt_cnm}.}
  \label{fig:aa_cent_eks_band}
\end{minipage}
\hspace{0.2cm}
\hfill
\begin{minipage}[t]{0.38\linewidth}
  \includegraphics[width=0.9\textwidth,clip=]{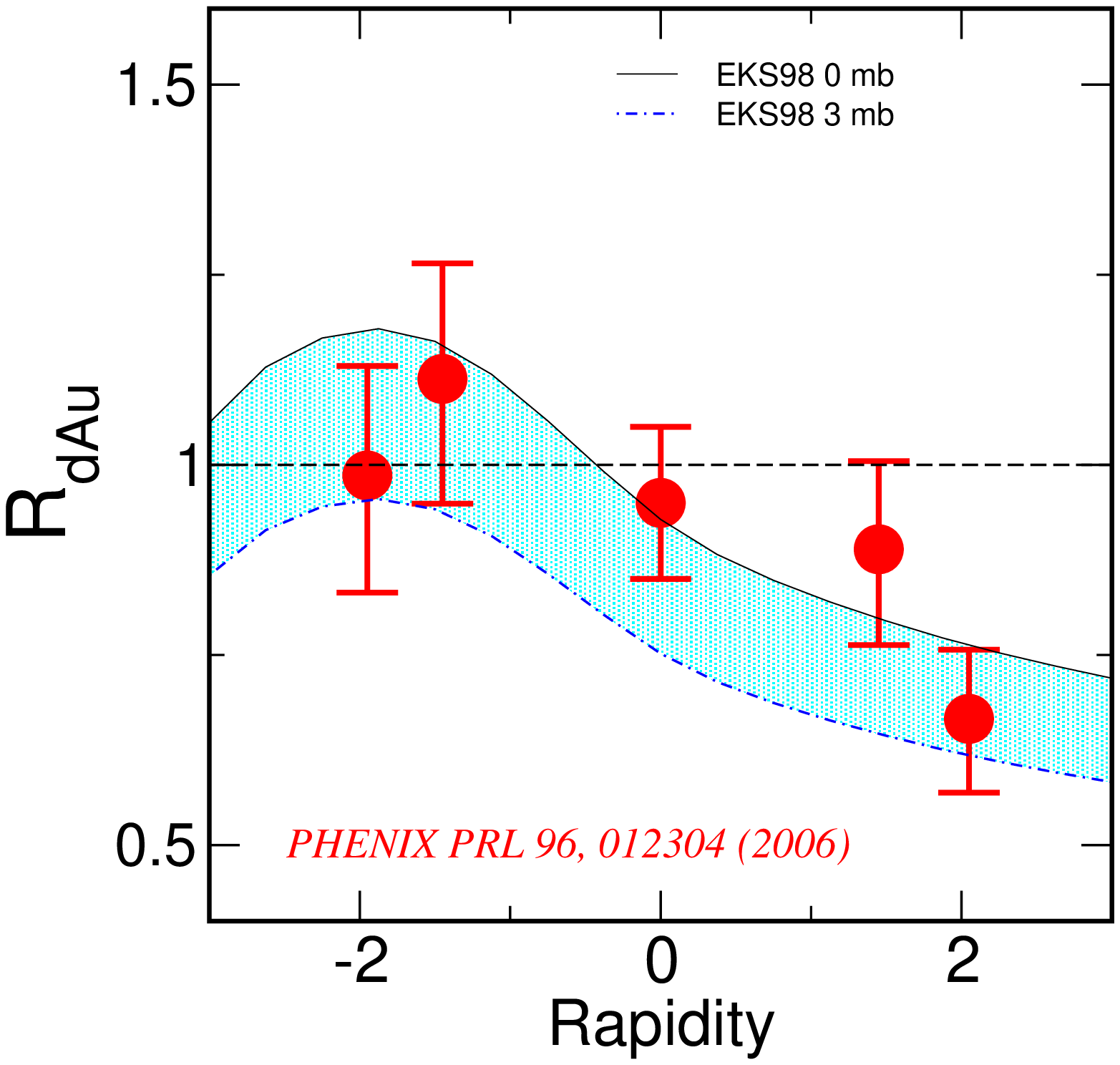}
  \caption{Results for $J/\psi$ suppression in d+Au collisions\cite{phenix_jpsi_pp} compared to a theoretical calculation
  that includes absorption and EKS shadowing\cite{vogt_cnm}.}
  \label{fig:rdau_eks98_band}
\end{minipage}
\hfill
\end{figure}

Final results from PHENIX for Au+Au collisions\cite{phenix_jpsi_run4auau} along with preliminary results
for Cu+Cu collisions\cite{phenix_qm05_jpsi} are now available, and are shown in Fig.~\ref{fig:aa_cent_eks_band}.
First it is important to understand what normal CNM $J/\psi$ suppression contributes in these A+A
collisions. This is illustrated by the blue error bands for A+A collisions in the figure
which represent theoretical calculations identical to those for the analogous blue error band in
Fig.~\ref{fig:rdau_eks98_band} for d+Au collisions. As can be seen, although the present d+Au data
lack the precision to constrain the CNM effects very well, there is still a clear suppression beyond
CNM effects in Au+Au collisions particularly for the forward rapidity data (blue points) and for the
most central mid-rapidity data (red points). Note that these CNM calculations probably do not explore
all possibilities for the resulting effects on A+A collisions, since they tend to be flat with rapidity
due to the approximate cancelation at forward rapidity of the shadowing of the small-x gluon and
the anti-shadowing of the other gluon. For example, gluon saturation may not provide
this cancelation. For Cu+Cu collisions the deviations below the CNM expectations are less clear.

\begin{figure}[tbh]
\hfill
\begin{minipage}[t]{0.48\linewidth}
  \includegraphics[width=\textwidth,clip=]{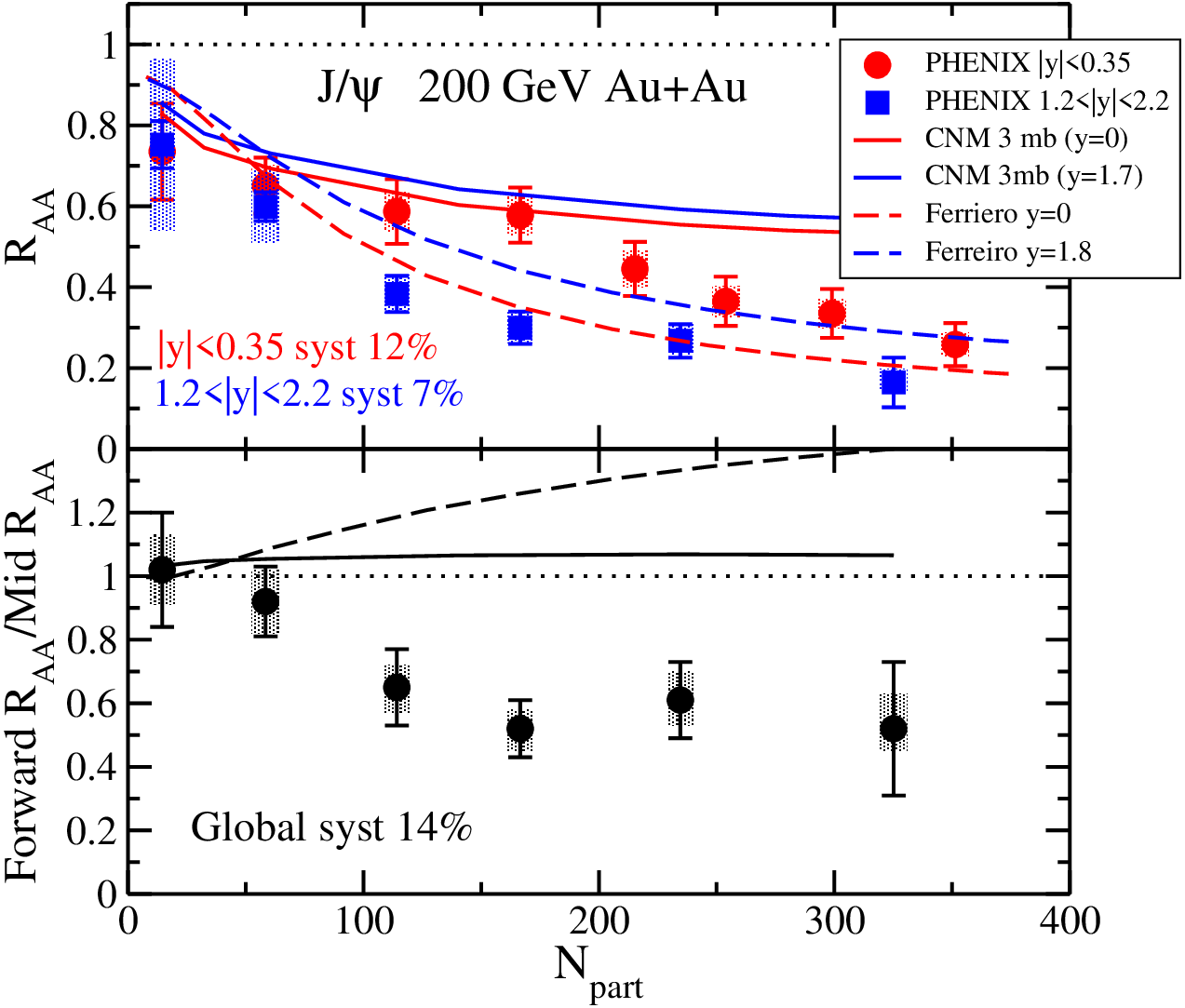}
  \caption{Nuclear modification factor, $R_{AA}$, for $J/\psi$ production in 200 GeV/c Au+Au
  collisions\cite{phenix_jpsi_run4auau}
  vs centrality (number of participants) in the top panel for mid (red) and forward (blue) rapdity. In
  the bottom panel the ratio of the forward over mid rapidty $R_{AA}$'s is shown.}
  \label{fig:raa_ratio}
\end{minipage}
\hspace{0.2cm}
\hfill
\begin{minipage}[t]{0.48\linewidth}
  \includegraphics[width=\textwidth,clip=]{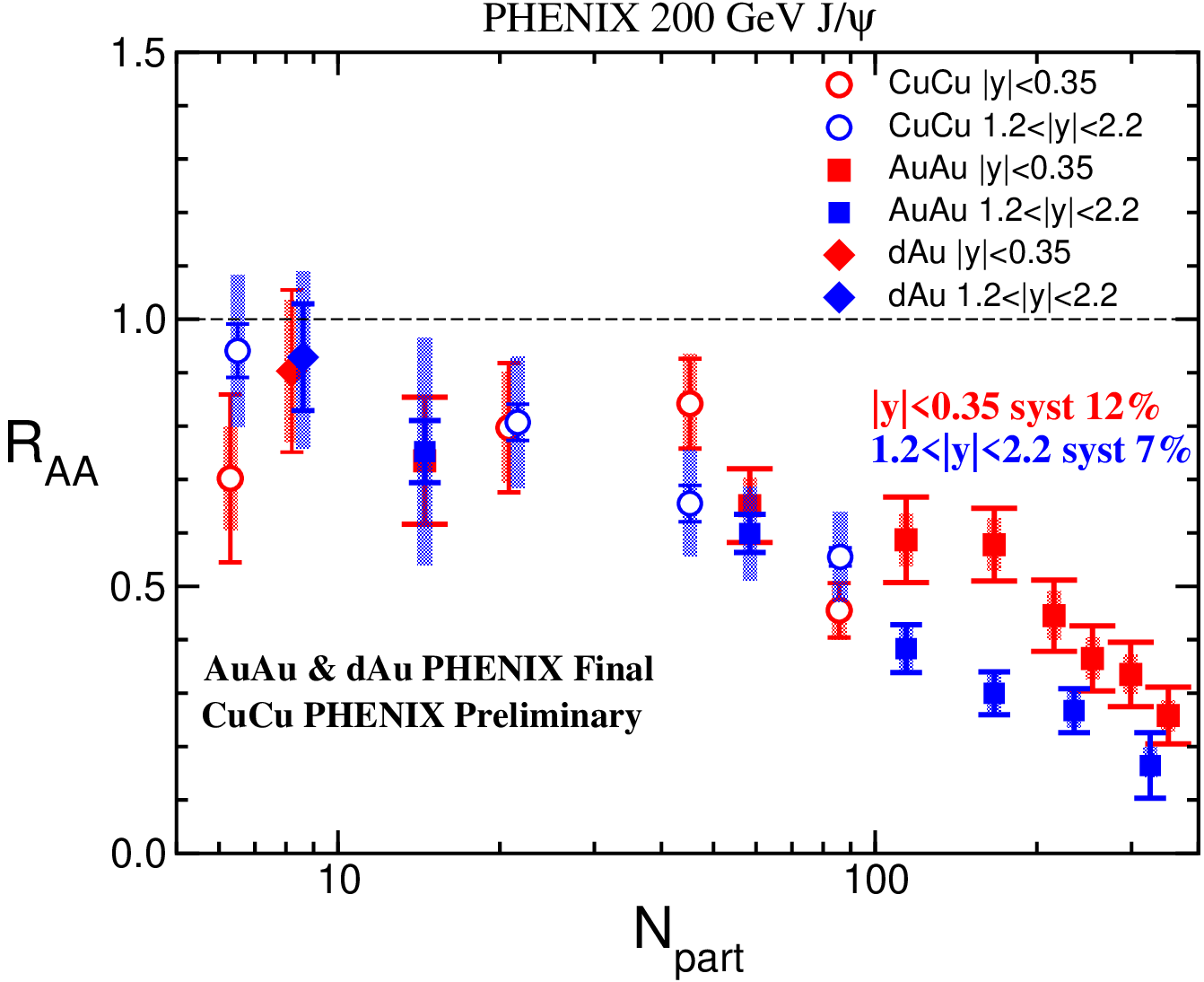}
  \caption{Comparision of final Au+Au results\cite{phenix_jpsi_run4auau}
  to preliminary Cu+Cu results\cite{phenix_qm05_jpsi} for the nuclear
  modification factor vs number of participants.}
  \label{fig:rboth_npart_data}
\end{minipage}
\hfill
\end{figure}

Looking just at the Au+Au data in Fig.~\ref{fig:raa_ratio}, one can see (top panel) that
the suppression for forward rapidity is significantly stronger than that for mid rapidity. In the bottom
panel the ratio of the $R_{AA}$ for forward rapidity to that for mid rapidity is shown,
and here the stronger forward rapidity suppression is quite distinct and this ratio appears to
show a saturation, within the experimental uncertainties, at about 0.6 for centralities above
$N_{part} \sim 100$. The features of
this ratio are undoubtably the most interesting from the new data and
will challenge theoretical interpretations. Also, as shown
in Fig.~\ref{fig:rboth_npart_data}, we see that the Cu+Cu results agree well with the Au+Au
data at small values of $N_{part}$ where they overlap.

\begin{figure}[tbh]
\hfill
\begin{minipage}[t]{0.48\linewidth}
  \includegraphics[width=\textwidth,clip=]{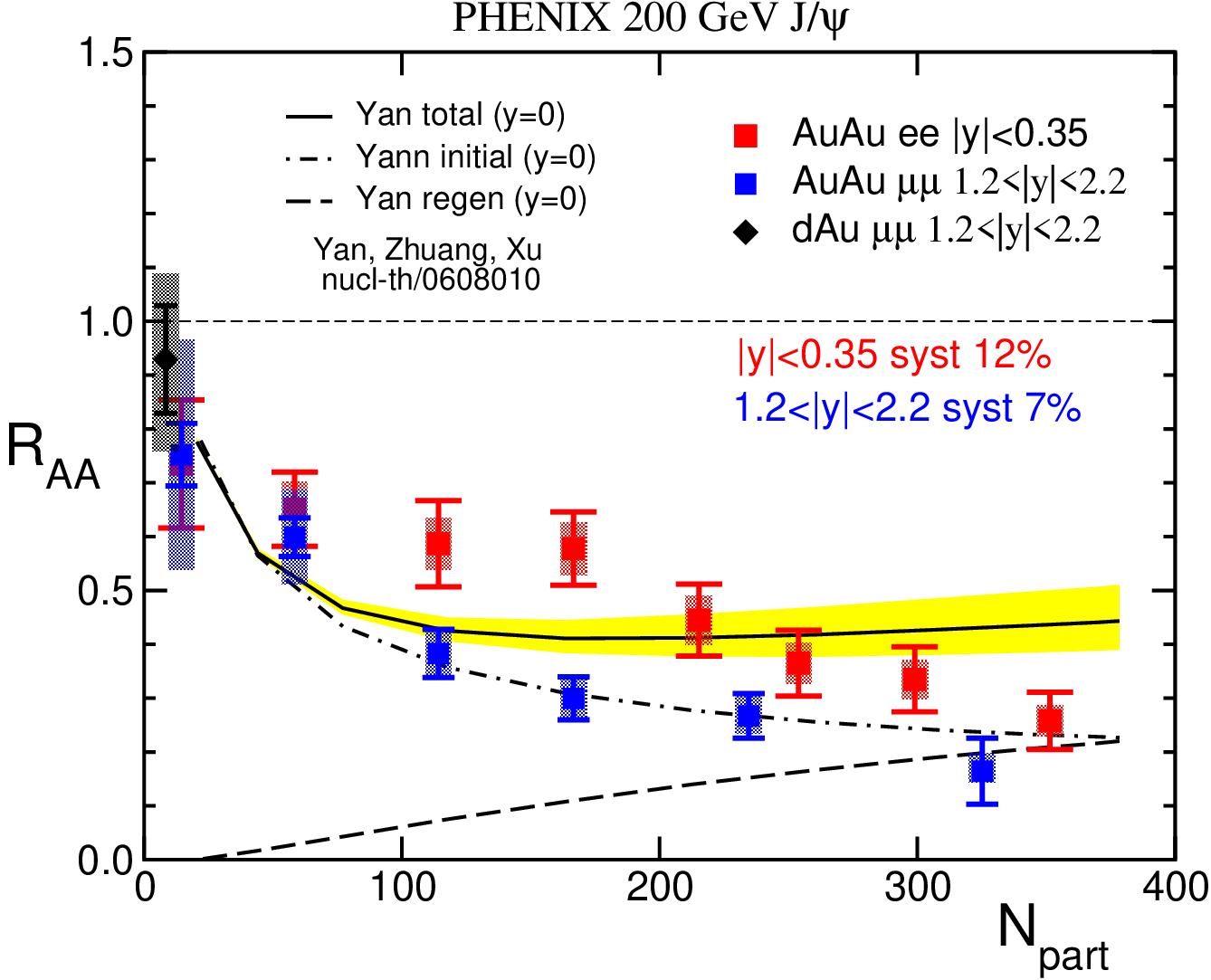}
  \caption{$J/\psi$ nuclear modification factor ($R_{AA}$) for Au+Au collisions at 200 GeV/c
  vs centrality (number of participants) for mid
  rapidity (red) and forward rapidity(blue) compared to theoretical calculations that include
  recombination\cite{yan}.}
  \label{fig:raa_npart_xu}
\end{minipage}
\hspace{0.2cm}
\hfill
\begin{minipage}[t]{0.48\linewidth}
  \includegraphics[width=\textwidth,clip=]{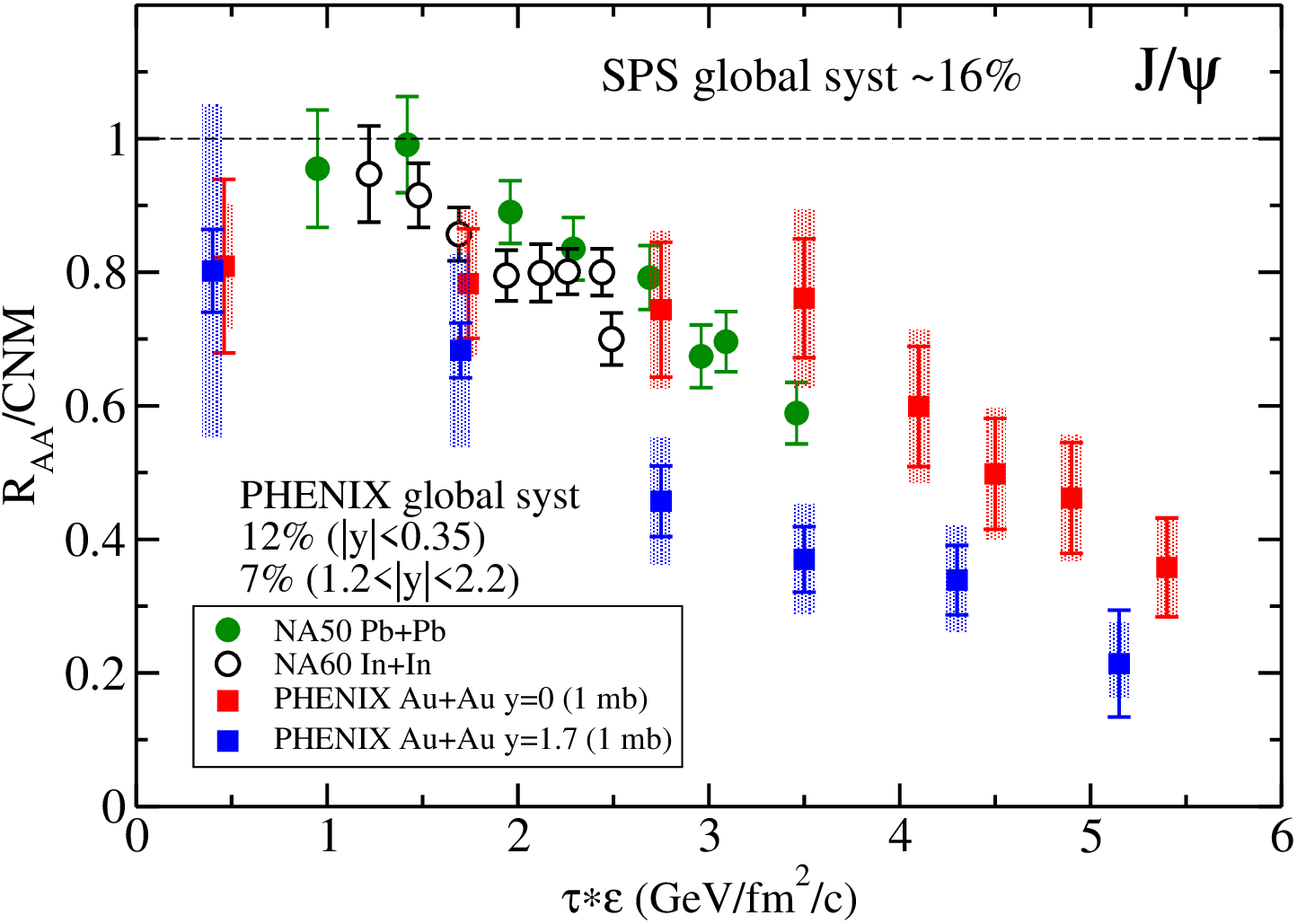}
  \caption{Survival fraction ($R_{AA}/CNM$) vs energy density comparison of PHENIX Au+Au
  suppression to that from NA38/50 at CERN.}
  \label{fig:saa_energy}
\end{minipage}
\hfill
\end{figure}

Numerous theoretical models\cite{capella, satz,rapp} were successful in describing the lower
energy SPS data, but all over-predict the suppression compared to the preliminary mid-rapditiy data at RHIC -
unless a regeneration mechanism is added as was done by Rapp\cite{rapp} and by Thews\cite{thews}.
The regeneration models provide an additional production mechanism for $J/\psi$s, where if the
total production of charm is high enough then charm densities in the final state will be sufficient
to give substantial formation of $J/\psi$s from coalesence of the large number of independent charm quarks created
in the collision. This production mechanism is predicted to be almost insignificant at SPS energies
but at RHIC may be substantial. This leads to a scenario in which strong screening or dissociation by a
very high-density gluon density occurs to a level of suppression stronger than that observed in the
RHIC data, but the regeneration mechanism compensates for this and brings the net
suppression back up to where the data lies. One of the recent calculations\cite{yan,rapp,thews} of this type
is shown in Fig.~\ref{fig:raa_npart_xu}.

In the regeneration picture, the stronger suppression at forward rapidity would result from the
lower density of charm at forward rapidity, which may be small enough to give no substantial regeneration
there. In this case the forward rapidity suppression would reflect the stronger suppression
from the QGP expected at RHIC compared to the
SPS. While at mid rapidity the higher charm density would provide substantial regeneration bringing the net
suppression back up to the same level as has been observed at the SPS. However both the screening
and the regeneration should increase with centrality, so the saturation in the forward/mid rapidity
suppression ratio challenges this picture.

An alternative interpretation of the preliminary results, sequential screening, is given
by Karsch, Kharzeev and Satz\cite{karsch}. In this picture, they assume that the $J/\psi$ is never
screened, as supported by recent Lattice QCD calculations for the $J/\psi$ - not at the SPS nor
at RHIC. Then the observed suppression comes from screening of the higher-mass states
alone ($\psi\prime$ and $\chi_C$) that, by their decay, normally provide $\sim$40\% of the observed $J/\psi$s. This
scenario is consistent with the apparently identical suppression patterns seen at
the SPS and RHIC for mid rapidity shown in Fig.~\ref{fig:saa_energy}.

However, the comparison shown in Fig.~\ref{fig:saa_energy} should be taken with caution, as it is not very clear
how to quantitatively compare the energy densities achieved at the SPS with those at RHIC.
Here we have used the Bjorken formula\cite{bjorken} with a $\tau_0 = 1~fm/c$ in both cases to estimate the
energy density, $\epsilon_{B_j} = {dE_T \over dy} {1 \over {\tau_0 \pi R^2}}$. Since the
crossing time at the SPS is about $1.6~fm/c$ it may be more realistic to use a larger $\tau_0$
there, while at RHIC the $\tau_0$ could be smaller than $1~fm/c$. The survival fraction ($R_{AA}/CNM$)
for the PHENIX points in this figure are obtained using CNM calculations like
those shown in Fig.~\ref{fig:aa_cent_eks_band},
with an absorption cross section of $1~mb$ and with uncertainties of $\pm 1~mb$ added into the
systematic uncertainties shown. For the SPS data we have estimated that an additional systematic
of about 17\% is appropriate - this is indicated on the figure but not added into the
SPS uncertainties shown.

In the sequential screening picture the stronger suppression at forward rapidity could come from
gluon saturation, which according to the CGC model\cite{cgc}, would not result in the flat rapidity
distributions obtained for CNM calculations like those shown in Fig.~\ref{fig:raa_ratio};
but instead would produce
a substantially smaller initial production at forward rapidities compared to mid rapidity.
However, this picture would also appear to be challenged by the saturation in the
forward over mid rapidity suppression ratio, since gluon saturation should continue to increase up
to the most central collisions.

\begin{figure}[tbh]
\hfill
\begin{minipage}[t]{0.48\linewidth}
  \includegraphics[width=\textwidth,clip=]{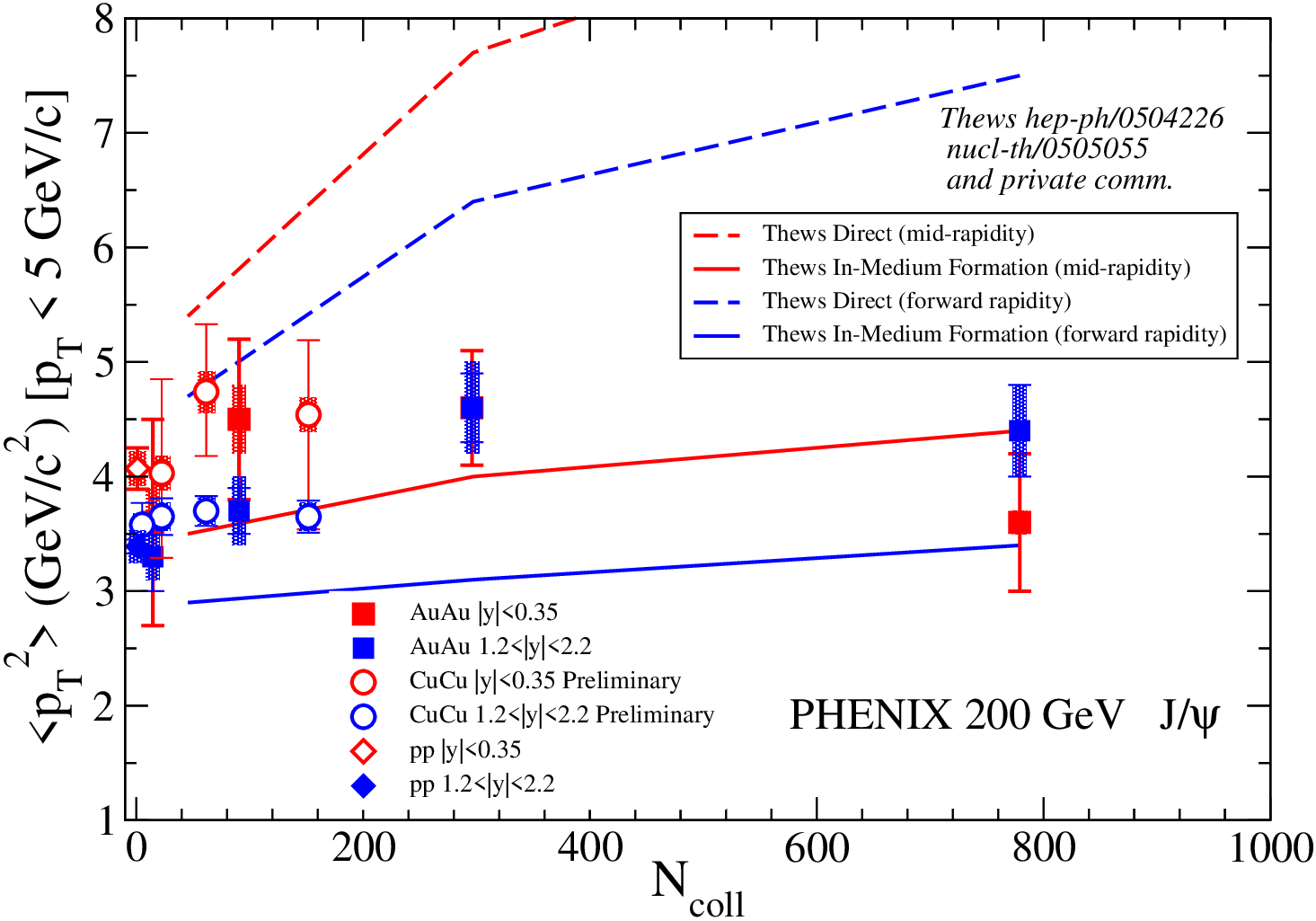}
  \caption{Mean $p_T^2$ vs centrality (number of collisions) of p+p, Cu+Cu and Au+Au
  data compared to model calculations\cite{thews}.}
  \label{fig:pt2_cucu_thews}
\end{minipage}
\hspace{0.2cm}
\hfill
\begin{minipage}[t]{0.48\linewidth}
  \includegraphics[width=\textwidth,clip=]{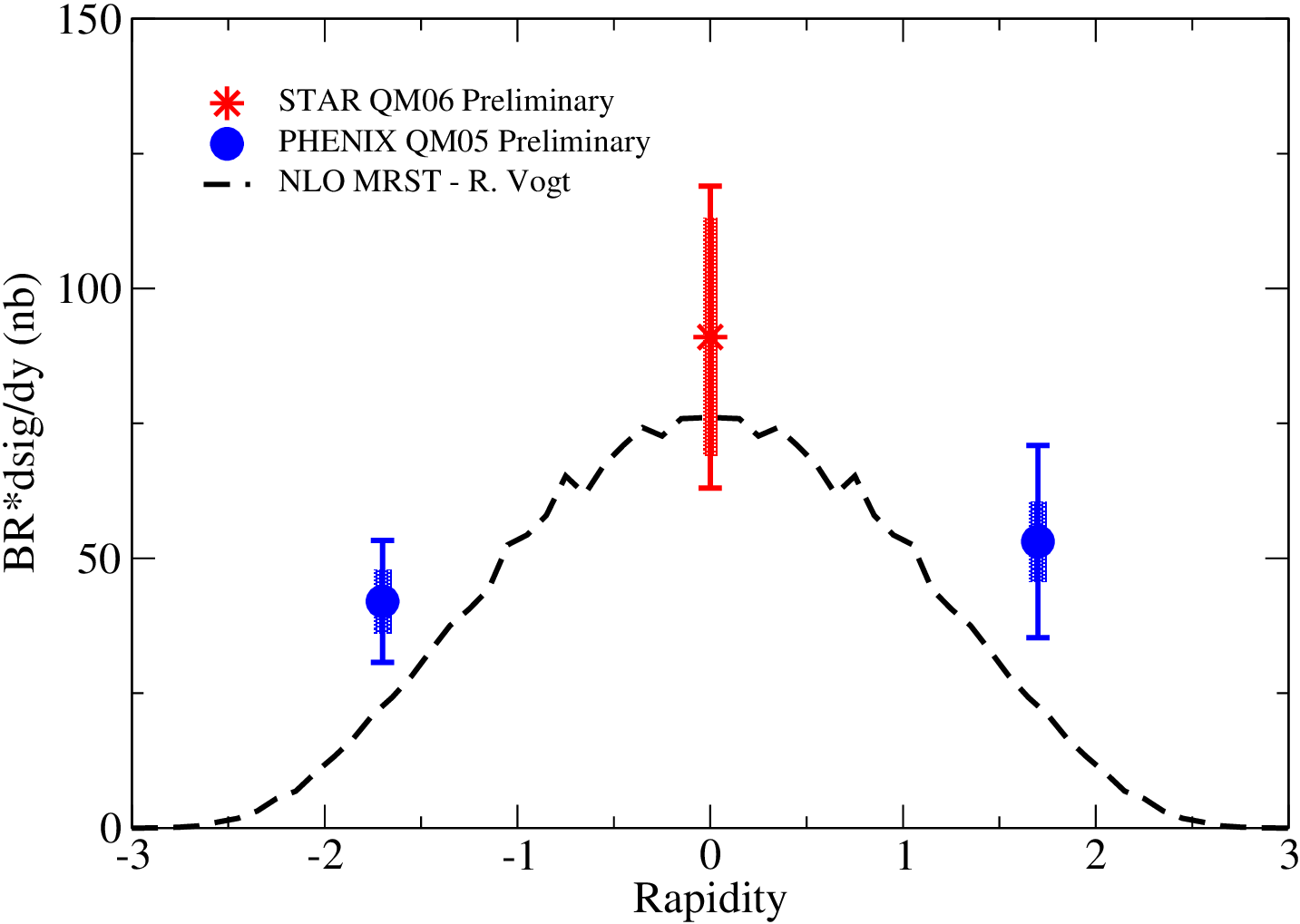}
  \caption{Preliminary measurements of the 200 GeV/c p+p cross section
  for $\Upsilon$ production from PHENIX\cite{PHENIX_ups} and STAR\cite{STAR_ups}.}
  \label{fig:ups_br_dsigdy}
\end{minipage}
\hfill
\end{figure}

Regeneration models also predict both narrowing of the $p_T$ distribution relative to the
usual Cronin broadening seen in p+A collisions, and, given that recent charm measurements
show flow, this flow should be inherited by the $J/\psi$s that are produced by regeneration.
Some evidence for narrowing of the $p_T$ has been observed - Fig.~\ref{fig:pt2_cucu_thews},
but a more reliable d+A CNM baseline will be necessary to establish this clearly.
Measurements of flow await the higher statistics of a new Au+Au run at RHIC.
As a result we are left with two different scenarios that provide
explanations for the RHIC A+A data. Both include the QGP in their picture, either through
color screening in the QGP or through dissociation of the $J/\psi$ by large gluon
densities.

Further advances in these important studies await higher luminosity runs for d+Au to solidify our
understanding of the baseline CNM effects, and higher luminosity A+A runs in order to capture sufficient
statistics for the rare quarkonia probes. A promise for the future can be seen in the recent $\Upsilon$ measurements
from PHENIX and STAR shown in  Fig.~\ref{fig:ups_br_dsigdy}.

\section{Summary}

Substantial uncertainties remain in the understanding of the production cross sections and the
polarization of charmonia. There are also a number of cold nuclear matter effects that influence
their production in nuclei and cloud our understanding of the suppression seen in nucleus-nucleus
collisions. Two competing pictures are able to explain the $J/\psi$ suppression seen in
nucleus-nucleus collisions at RHIC - one involving sequential screening in the plasma of the various
charmonia states; the other with strong dissociation of all charmonia states by a dense gluon field
but recombination of independently produced charm quarks. More precise measurements in the future will
be necessary to distinguish between these two quite different scenarios.

\end{document}